# Reading Protocol: Understanding what has been Read in Interactive Information Retrieval Tasks

Daniel Hienert*, Dagmar Kern*, Matthew Mitsui[+], Chirag Shah[+], Nicholas J. Belkin[+]

*GESIS – Leibniz Institute for the Social Sciences
Cologne, Germany
{firstname.lastname}@gesis.org

[+]School of Communication & Information
Rutgers University
New Brunswick, NJ, USA
{mmitsui, chirags, belkin}@rutgers.edu

## ABSTRACT

In Interactive Information Retrieval (IIR) experiments the user's gaze motion on web pages is often recorded with eye tracking. The data is used to analyze gaze behavior or to identify Areas of Interest (AOI) the user has looked at. So far, tools for analyzing eye tracking data have certain limitations in supporting the analysis of gaze behavior in IIR experiments. Experiments often consist of a huge number of different visited web pages. In existing analysis tools the data can only be analyzed in videos or images and AOIs for every single web page have to be specified by hand, in a very time consuming process. In this work, we propose the reading protocol software which breaks eye tracking data down to the textual level by considering the HTML structure of the web pages. This has a lot of advantages for the analyst. First and foremost, it can easily be identified on a large scale what has actually been viewed and read on the stimuli pages by the subjects. Second, the web page structure can be used to filter to AOIs. Third, gaze data of multiple users can be presented on the same page, and fourth, fixation times on text can be exported and further processed in other tools. We present the software, its validation, and example use cases with data from three existing IIR experiments.

## CCS CONCEPTS

• Information systems~Users and interactive retrieval

## 1 INTRODUCTION

A significant part of research in Interactive Information Retrieval (IIR) is to study user behavior in the context of search tasks. Therefore, in IIR experiments, the user's interaction with the system is recorded by capturing keyboard, mouse and browser actions, but also by recording the user's gaze motion on the monitor with eye tracking. This allows the analyst to understand which regions the user has looked at. In (I)IR, interesting findings between gaze behavior and concepts such as relevance [2, 18], user interest [1, 16], knowledge level [13] or task types [15] has been found.

Eye tracking software records the stimulus data (what is shown on the monitor) as images and videos with the benefit of simple data collection. However, the disadvantages become apparent later during data analysis: Gaze data can be shown and analyzed again only on images and videos, e.g. as gaze plots or heat maps. Thereby, the underlying website structure and the visible text can be accessed only with great effort, e.g. by manual inspection of the videos or by drawing areas of interest (AOI) manually on the stimuli images.

In this paper, we introduce the reading protocol tool. The software exploits the idea of mapping gaze data down to the textual level. As a result, it gives the analyst much more possibilities for analyzing eye tracking data. It can be directly seen over tens or hundreds of stimuli pages what has been viewed and read by the subjects on the text level. The whole data set can be immediately filtered down to certain participants, stimuli pages and areas of interest. Fixation times for all this data can be exported and further processed in other tools. In the context of IIR, this gives much more possibilities to understand what has actually been viewed and read by users within a search task.

In the following, we will present related work in the areas of IIR, eye tracking, and reading behavior. We will then present the reading protocol tool, its validation and will show its capabilities with data from three existing IIR experiments and will then discuss the pros and cons of the tool.

## 2 RELATED WORK

### 2.1 The Task in IIR

The *task* plays an important role in the IIR search process. The need to conduct a task arises from a *problematic situation* in which a user realizes that she lacks knowledge about a specific problem, topic or situation [3]. A search task is then the activity to accomplish the *goal* of receiving information about the specific issue [35]. An overall goal can be divided into several sub-goals which are processed by a user with different *information seeking strategies* (ISS) [4]. Models in IIR put these basic concepts into an overall system. As a turning point in IR research, the classical laboratory framework has been extended by Ingwersen & Järvelin [21] with the seeking-, work-task-, socio-organizational and cultural context. Borlund [9] proposes an IIR evaluation model with the help of the simulated work task. This can be used to evaluate IIR systems as realistically as possible, but also relatively controlled. Another framework is the usefulness evaluation model [12] which tries to measure the usefulness on the entire seeking episode, on each interaction, and on the system support level. In order to study different task characteristics, it can be differentiated between different *task types*. Kellar et al. [24], for example, differentiate between fact finding, information gathering, browsing and transactions. Li & Belkin [26] apply a faceted task classification system to describe a task on facets such as the source of task, task doer, time, action, product, and goal. Recently, in IIR, the role of *learning* in the search process has become more prominent [34]. This view tries to understand how the interaction with information leads to the modification of a searcher's knowledge structure.

## 2.2 Eye Tracking in IR

Eye tracking as a method to capture the user's gaze is used in different disciplines such a Marketing, Psychology or HCI. It is also used as a method to study user behavior in information search, e.g. for general web search [17] or over different task types [23]. On a more conceptual level, gaze data can also be used as a source for implicit user feedback. For example, a number of research works [e.g. 2, 18] try to find relationships between gaze behavior and the concept of *relevance*. Other research uses gaze data to identify the search *interest* [1] or the *intention* of the current search query [33]. Gaze data can also be used proactively for *re-ranking* and *query expansion* [10]. Specifically, in IIR eye tracking data has been used to predict the user's knowledge level [13] or to discriminate between different task types [15], also in relation to task facets [14].

The idea of mapping eye tracking data to the Document Object Model (DOM) of a web page and to the web page text has already been applied in earlier systems. After an eye tracking experiment, WebEyeMapper and WebLogger [27] can be used to map eye tracking data and cached web pages to specific HTML elements and their text content. The results are stored in a database. WebGazeAnalyzer [5] uses the same basic process but also allows to analyzes reading behavior on web pages with metrics such as reading coverage, regressions and reading speed. Unfortunately, these earlier research systems seem to be discontinued and not publically available. More recent research uses eye tracking data on the word level e.g. for the analysis of topical interests [16] or for showing multimedia content based on the read word [6].

Gaze data can be visualized with a number of visualizations such as gaze plots or heat maps, a recent overview of visualization types is given in [7]. However, most visualization techniques are focused on showing the overall gaze behavior on the whole stimulus page. More specialized visualization techniques for reading behavior are presented in applications such as EyeMap [32] or GazePlot [31] which show fixations and saccades at the word level.

## 2.3 Reading Behavior

A common model for the human reading process is the E-Z reader model [28]. This model assumes that reading is a serial process in which the reader fixates one word at a time (fixation) and then shifts to the next word (saccade). The fixation is divided into two stages: L1 - the basic word identification ("familiarity check") and L2 understanding the meaning ("lexical access"). In reading tasks fixation times can be e.g. 122ms for L1 and between 151ms and 233ms for L1+L2 depending on word length, word frequency and the word/text difficulty [28]. In non-reading tasks, e.g. the visual search of target words, fixation times can differ [29]. Reading behavior can also differ much in real-world settings [22] on the different levels of reading words and sentences, whole text comprehension and the integration over multiple documents. Reading behaviour is also influenced by the reader with different expertise, knowledge, attitude, ability and especially the task one is doing.

# 3 THE READING PROTOCOL TOOL

In this section, we present the reading protocol tool. We first outline the main concept, give an overview of the user interface, describe the needed input data and explain briefly the computation of word-eye-fixations.

### 3.1 Main Idea

Eye tracking data in all common analysis software is most often visualized as (gaze)-videos, heat maps or gaze plots ([7] gives an overview). This makes the analysis process costly because video- and image-based data needs to be inspected manually by the data analyst. The connection between gaze data and underlying data such as the website structure or the text is lost and cannot be further processed. The reading protocol tool maps eye tracking data down to the word level of the website's textual content. In reading protocol's user interface the gaze data is then shown on the textual level what gives the analyst much more possibilities for the analysis.

### 3.2 User Interface

The user interface of reading protocol mainly consists of three different components: (1) the filter menu, (2) the stimulus section and (3) the overall data table. An instance of reading protocol can be found under the address www.vizgr.org/reading_protocol.

The filter menu (Fig. 1a) contains a number of filters with which the analyst can filter down the data set of the whole IIR experiment. There are filters for subjects, stimuli pages, and AOIs described by Cascading Style Sheets (CSS) labels. The filtered data set is then shown accordingly in the stimulus section and in the overall data table. There are three sliders to control the coloring behavior of fixated words and for hiding non-fixated text sections.

The stimulus section (Fig. 1b) shows per default for every subject each stimulus in chronological order. Words which have been fixated are coded along a cold (blue) to hot (red) color scale. Words and text passages which has been viewed longer (e.g. 122ms and above) can be identified easily by their more hot-colored backgrounds. Words and text passages that have been viewed only by scanning (e.g. everything under 122ms) are only slightly cold color-coded in violet.

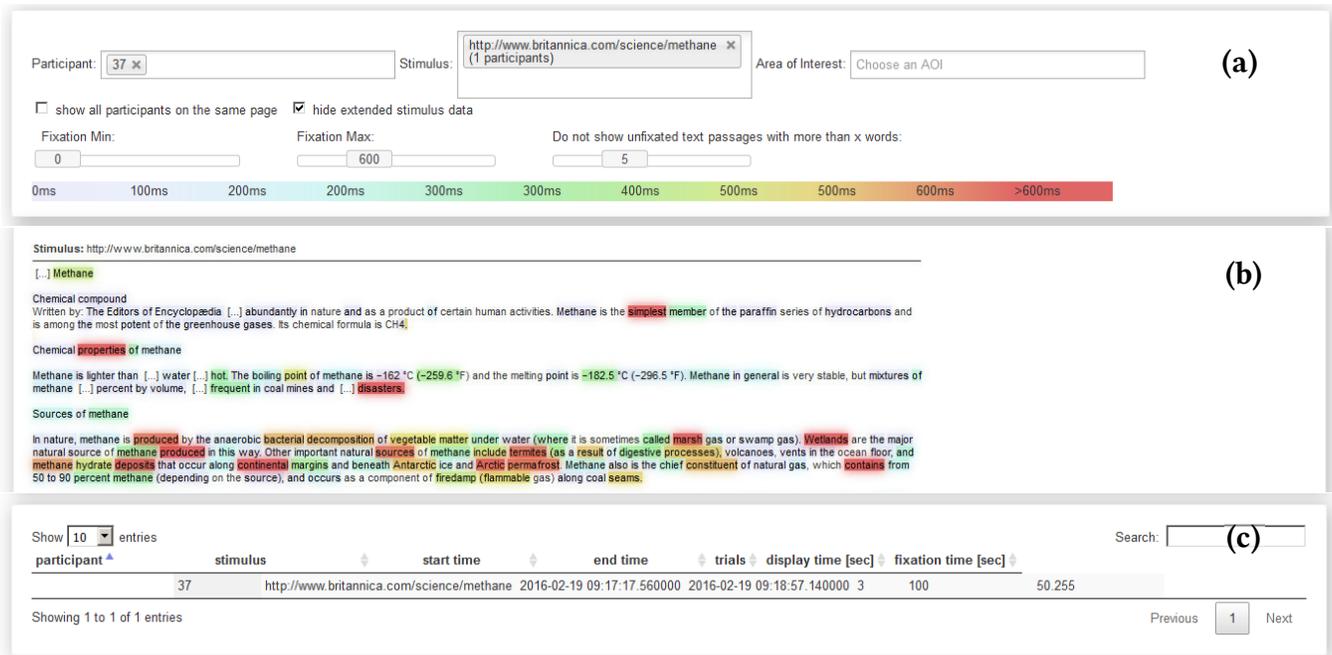

Figure 1: Reading protocol's user interface

The data table (Fig. 1c) then contains all information in a table format which can be either sorted and searched or copied to external tools.

### 3.3 Input Data

Reading protocol needs two input data sets: (1) raw eye gaze data from the eye tracking software. This can normally be exported as CSV from the eye tracking software. (2) the stimuli pages as HTML. These can either (a) be recorded live in new experiments with a number of browser plugins, (b) be taken from existing experiments if recorded or (c) be transformed from existing stimuli images of existing experiments via OCR. Compare the reading protocol software page for details[1]. Reading protocol then processes this data and builds the word-eye-fixations object. This is a JSON structure which contains every word from the stimulus page that has been viewed by the participant identified by the eye tracking x, y coordinates. For every word, there is the word position (based on the web page as text), the aggregated fixation time, and a timestamp for the first and last time the user has viewed this word.

### 3.4 Computing Word-Eye-Fixations

To compute eye fixations on words, the algorithm first opens the stimuli web page in a browser in the original layout. Raw gaze coordinates are loaded from the database. Fixations and saccades are separated with an I-DT algorithm [30]. For each fixation coordinate we use a browser-specific method (e.g. *caretPositionFromPoint* in Firefox) to determine the word under these coordinates. Fixation times, time stamps, word position and other information are collected for each individual word over all coordinates and are later saved in the database. For more details on the algorithm check the code in the repository[1].

## 4 EXAMPLE DATASETS

To demonstrate the tool's functionality in the following section, we use three existing datasets from IIR experiments. Two come from domain-specific literature search in the social sciences, and one dataset derives from web search in the area of journalism.

(1) The first dataset comes from a lab study with two groups of 16 subjects each. All subjects worked in different fields of the social sciences. The first group consisted of bachelor and master students; the other group was built from postdoctoral researchers. The students were between 22 and 35 years old (m=26.38; 12 female, 4 male) and the postdocs were between 30 and 62 years old (m=40.19; 8 female, 8 male). All subjects were recruited via e-mail and personal recommendation. All subjects were given the same document about "education inequality". Their task was to find similar documents in the social science literature portal Sowiport [20]. They were free to use different search strategies like author or keyword search, and they had ten minutes time to solve the task. The main research question was to find out which search strategies users would apply.

The subjects used a keyboard, a mouse and a 22″-monitor connected to a laptop. The laptop display served as an observation screen for the interviewer who was with the subject in the same room during the study. For tracking their eye movements, a remote eye tracking device SMI iView RED 250 was used with a 0.4 degree gaze position accuracy. It has been attached to the bottom side of the stimulus monitor. The eye tracker was calibrated with each subject using a 9-point calibration with a sampling frequency of 250Hz. More details on the experiment and the results can be found in [11]. We refer to this experiment as the "search strategies"-experiment.

---
[1] The reading protocol software is open source and can be found under https://git.gesis.org/iir/reading-protocol

Eye tracking data for this experiment was exported with the SMI BeGaze tool as CSV and imported into the reading protocol database (~3 million rows). Subjects viewed altogether 511 detailed record pages which were downloaded from the Sowiport portal including all JS and CSS and stored in the reading protocol database.

(2) The second dataset consists of data recorded during another lab study with the same literature portal Sowiport but with different tasks and subjects. The 25 researchers in the field of social science (16 female, 9 male, age ranging from 23 to 45 years, 14 held a bachelor degree, 10 a master degree and one is a postdoctoral researcher) were recruited through the IIRpanel[2], mailing lists, and posters at the local university. Participants were asked in simulated work task scenarios to search for and to bookmark relevant publications either to a topic they are familiar with or to an unfamiliar topic. The topic was individually chosen by the participants themselves. In this study, two different conditions of representing an abstract were tested. In one case, the 20% most important keywords (based on tf-idf calculations) were highlighted in yellow. The other case served as a baseline with no highlighted words. With this study, we wanted to find out how abstract with highlighted words affect the information search behavior.

In single sessions, participants set in front of a 22″ monitor, connected to a laptop that ran the eye tracking software form SMI and used keyboard and mouse to perform their tasks. The eye gazes were recorded by remote eye tracking device SMI iView RED 250 using a sampling rate of 60Hz. The interviewer was observing the screen and gaze activities in a different room. Results on this experiment can be found in [25]. We refer to this experiment as the "highlighting"-experiment. Here again, eye tracking data was transferred from the SMI BeGaze tool to reading protocol via CSV (~2.6 million rows). Subjects looked at 1,272 detailed record pages which have been downloaded from Sowiport with JS and CSS and stored in the reading protocol database.

(3) For the third experiment, we use study data from a lab study with 40 undergraduate students from undergraduate journalism courses having completed at least one course in news writing. Subjects had to perform two different search tasks on two different search topics. For this work, we use only data for the topic: "Methane Clathrates and global warming". There were four different search tasks to choose from, but in this experiment, we only use two. The first task was Copy Editing (CPE): subjects had 20 minutes time to check the accuracy of six italicized statements from a given text. They should find and save web pages that confirm or disconfirm the statements. For this task, we have valid eye tracking data for 9 subjects (from 19 years to 21 years; m=20.33; 7 female, 2 male). The second task was Story Pitch (SP): subjects should find and save web pages that contain the six most interesting facts about the world economic impact of global warming on the Arctic Region. Here, we have valid eye tracking data from 7 subjects (from 20 years to 23 years; m=21.14; 4 female, 3 male). Both tasks have the same topic, but can be characterized by different task facets [26] as shown in table 1.

Table 1: Task facets in the journalism experiment

| Task Name | Task Facets | | | |
|---|---|---|---|---|
| | Product | Level | Goal | Named Items? |
| Copy Editing (CPE) | Find facts | Segment | Specific | Yes |
| Story Pitch (SP) | Find facts | Segment | Amorphous | No |

Subjects here used a keyboard, a mouse and a 24″ monitor (1920x1080 pixels). Their activity was recorded with the Firefox browser plugin Coagmento and Morae[3]. Eye tracking was conducted with a GazePoint GP3 tracker, a 60 Hz bright pupil tracker with 0.5-1 degree accuracy. Participants were calibrated using 9-point calibration. Details on the experiment design can be found in [19]. We refer to this experiment as the "journalism"-experiment.

The eye tracking data for this experiment was exported from the GazePoint Analysis software to CSV. Because here, the eyes' coordinates were stored absolute to the monitor, before importing, we added the scrolling offset and additionally the browser's header offset. For CPE (Copy Editing), we have around 850,000 rows, for SP (Story Pitch) around 550,000 rows. Within the experiment, the pure HTML code of the viewed web pages was saved by Coagmento. Before importing to reading protocol, the corresponding JS and CSS files were crawled from the Wayback Machine[4]. All pages were then manually verified with the gaze videos for their original layout as in the experiment. For CPE we have 73 content pages, for task SP we have 70 pages.

## 5 Tool Validation

Building the word-eye-fixations object has a number of issues which needs to be addressed and validated. Eye tracking coordinates can be either absolute to monitor or relative to the browser's viewport. In the first case, the user's scrolling behavior needs to be taken into account. The stimulus page has to be in the original layout as it has been shown in the experiment when building word-eye-fixations from existing experiments. As a first attempt, we compared visually heat and gaze maps from existing software to the ones created with reading protocol. Figure 2 shows an example from the highlighting experiment: (a) a heat map of the stimulus page from the original eye tracking software (SMI BeGaze), (b) rendered eye-tracking points with the reading protocol script, (c) the textual view in reading protocol.

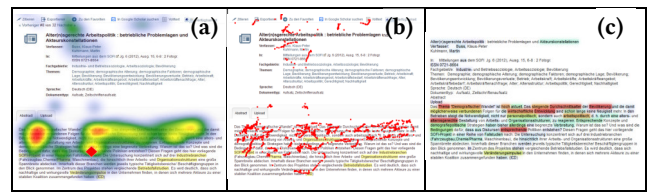

Figure 2: Comparing eye tracking coordinates: (a) a heat map from SMI BeGaze, (b) rendered eye-tracking points with reading protocol, (c) the textual view in reading protocol

---

[2] https://multiweb.gesis.org/iirpanel/?en
[3] https://www.techsmith.de/morae.html
[4] https://archive.org/web/

A second validation was done by comparing fixation times in AOIs of the same dataset between BeGaze and the reading protocol. In the highlighting experiment (cp. section 4.2) subjects had to search and bookmark documents to a topic they are either familiar or unfamiliar with. In BeGaze we manually drew AOIs over all record sections (the document's metadata + abstract) and abstract sections in n=990 stimuli and exported the AOI fixation times. Both AOIs differ in size and the abstract AOI also differs in its position on the page. Similarly, in reading protocol we filtered all stimuli pages to the record and the abstract section and exported fixation times. Note that the algorithm in BeGaze for computing fixations differs from the one in reading protocol: in reading protocol we only count eye tracking coordinates over words, not in white spaces. However, by comparing the data, we found a correlation between the two approaches. Table 2 shows the exact results. As expected, the mean fixation times of reading protocol (RP) are a bit under those of BeGaze, because of the fact that the reading protocol does not consider whitespaces. We found a very strong Pearson correlation between both data rows of 0.912 for the record AOI and of 0.957 for the abstract AOI (both with $p<0.0001$). Figure 3 shows the graph of fixation times over all stimuli pages. It can be seen that the reading protocol curve (blue in Fig. 3) tracks the BeGaze curve (red in Fig 3) accurately.

**Table 2: Fixation times in sec. comparison between data analyzed with BeGaze and with the Reading Protocol (RP)**

| AOI | Tool | Max | Mean | SD | Pearson |
|---|---|---|---|---|---|
| Record | BeGaze | 113.245 | 13.873 | 15.399 | 0.912 |
| Record | RP | 92.446 | 11.468 | 13.149 | 0.912 |
| Abstract | BeGaze | 95.032 | 9.248 | 13.346 | 0.957 |
| Abstract | RP | 87.070 | 7.730 | 11.792 | 0.957 |

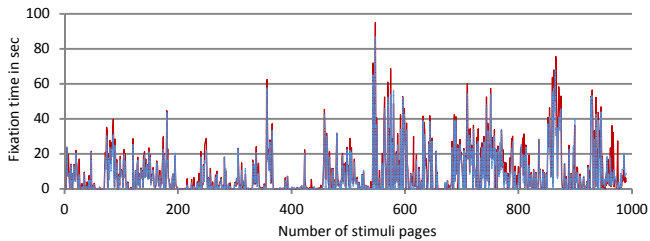

**Figure 3: Fixation times in sec over all stimuli for the AOI Abstract. Reading Protocol is in blue and BeGaze in red.**

## 6  USAGE OF THE READING PROTOCOL

In this section, we show how the reading protocol tool can be used to analyze data sets from IIR experiments. We will show the functionality and application with the help of the example datasets described above. We do not want to give a full-scale analysis on reading behavior in web search here but we want to give examples how reading protocol can support analyzing data from different experiments.

### 6.1  From Gaze Images/Videos to the Text Level

Existing eye tracking analysis software provide a number of visualization techniques to analyze gaze data (cp. [7]). For example, the user's scan path can be replayed as an overlay on stimulus images or videos. Or the fixations are shown as a heat map over the stimulus image. In common for all different visualizations is that the gaze data is shown over stimulus images/videos to let the analyst understand where exactly and for how long a subject has looked at on a stimulus. However, the underlying text can only inspected manually – so the analyst has to check which regions are hot-colored and has then to extract manually from the image/video what is the text underneath.

In reading protocol, we transfer from stimulus images and videos to the textual level. This has some advantages which are shown in this and the next paragraphs. However, the core idea is to understand what has been viewed and read on the textual level in the context of an IIR task.

For every stimulus page, gaze data is encoded as colored backgrounds of the words in the sense of a heat map. Words with only slight fixations times (0 to 100ms) are colored in light-violet, words with higher fixations times (>100ms) are colored from blue over green/yellow/orange to red. This way, the analyst can instantly see which stimuli pages have only be scanned and which passages have been viewed and read more intensively.

To allow an even better overview on the stimuli level, we use two techniques: (1) with two sliders the analyst can control the coloring behavior. "Fixation min" sets the starting point, "fixation max" the end point of the color scale. For example, one can color only words with more than 122ms fixation time ("lexical access") and above. (2) The analyst can control with another slider how many consecutive words are hidden that have not been fixated at all. This hides larger text passages from the page which have not been viewed and let the stimuli collapse. Figure 4 shows a stimulus page from the journalism task Story Pitch where the subject has inspected the Wikipedia page for "Methane Clathrates".

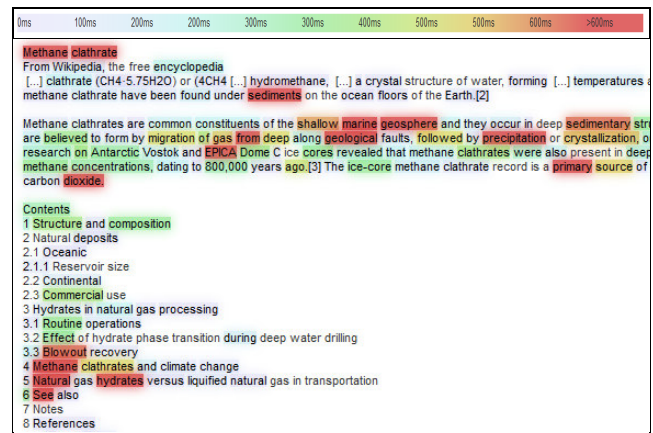

**Figure 4: A viewed Wikipedia article from the journalism task Story Pitch. The first paragraph has been only scanned; parts of the first paragraph are collapsed in "[…]". The second paragraph has been extensively read. Note, how the subject is also interested in the "Methane clathrates and climate change" line in the list of contents.**

### 6.2  Task-based experiments

The task is an important concept in IIR (see Section 2.1) and also the starting point of many experiments conducted in IIR. A lot of data is collected within such experiments such as log data with user actions, keyboard and mouse data, questionnaires, but also eye tracking data. Often research in IIR tries to find correlations between user behavior and recorded data. Existing eye tracking analysis software take into account the concept of the task only by labeling the task type to subjects. However, for further analysis, visualizations techniques from the analysis software have to be used or raw gaze data has to be exported and further processed in other statistical software.

Reading protocol supports the concept of the task in different ways. First, stimuli pages in reading protocol are arranged in chronological order in the task so that an analyst can see which pages have been viewed and which text has been viewed and read by the user. All stimulus pages are shown on the same user interface and can be overviewed by scrolling. If the amount of data is too much, it can be filtered by participant, stimuli and additionally by areas of interest. This allows a quick overview of the data and the reading behavior. Second, task-related data and data from other resources can be additionally shown per stimulus and in the data table. For example, one can add information for the task type (A or B), the task topic, the perceived task difficulty from the questionnaire and information such as the stimuli's usefulness rated by the user. Figure 5 shows a stimulus page from the highlighting experiment with extensive annotated task and stimulus data. This example also shows that an analyst can easily see whether correlations exist between the words read and the additional displayed annotated task and stimulus data. One can see, e.g., that words with longer fixation times refer to the topic on which the participant was looking for, in this case, "well-being of migrants".

Figure 5: Part of a stimulus page from the highlighting experiment with extensive annotated task and stimulus data on the right side.

### 6.3 Areas of Interest

Existing eye tracking software holds stimulus data as images. Each stimulus which has been viewed by a subject (e.g. a web page) is saved as a separate screenshot. Often analysts are only interested in certain parts of the stimulus because the experiment focuses on it. For example, in literature search one can ask what the influence of the title vs. the abstract section is.

In existing analysis software areas of interest have to be created manually by the analyst as stimulus data are images. AOIs have to be created manually for each stimulus. Therefore, the analyst has to draw the AOIs on the stimulus image with forms such as rectangles, polygons or ellipses. This has to be done for each stimuli-subject combination. In large experiments with hundreds and thousands of stimuli pages, this is very time-consuming. Additionally, AOIs can change over time, for example, when websites are changing their appearance with dynamic elements.

In reading protocol the analyst can filter all stimuli pages to certain AOIs with a mouse click. Within the rendering process reading protocol extracts all CSS IDs and class information of the stimuli pages. In the user interface, the analyst can choose one or more labels from the filter menu and all stimuli pages are filtered to these elements. Additionally, the fixation time, number of words and characters fixated and the percentage of words fixated for the chosen AOIs are shown in the page information and in the data table.

*Example Search Strategies:* In the search strategies experiment it is interesting to understand for the analyst which document sections such as title, authors, keywords etc. has been viewed and which has been used for browsing in the document collection in the sense of the task. For users, there are several possibilities, e.g. browsing by author names, category, keywords, related entries, references, citations etc. The analyst can filter down to these areas of interest, can view and compare reading times. Figure 6 shows stimuli pages from this experiment showing only the AOIs title, authors, source, category, and keywords.

Figure 6: Showing three stimuli from the search strategies experiment. Only the AOIs title, authors, source, category and keywords are shown, the rest is hidden. This user is inspecting document keywords very intensively.

### 6.4 Multiple users

In existing analysis software, it is possible to integrate eye tracking data of multiple users. For example, in SMI BeGaze a heat map visualization can show the accumulated data of multiple users. Fixations for all users are summed up and shown on the same stimulus page. With this, the analyst can identify what are hot regions on the same stimulus page over all subjects.

However, this approach has certain limitations: first, data on a stimulus can only be aggregated if the stimulus is exactly the same. But, web pages' layout can change because of dynamic elements such as news tickers or advertisements. Also, if experiments are conducted over a longer time period, web pages can slightly change, e.g. because of a new design. In such cases, for each layout version, a new stimulus image is created and the data of different users cannot be shown anymore on the same stimulus page. Second, eye tracking data is again only shown on images. So, stimuli pages cannot be filtered down to certain regions, fixation data for these regions are missing, and viewed text cannot be extracted and further processed.

**Table 3: Statements extracted from the task description which has to be confirmed in the Copy Editing task of the journalism experiment**

| | |
|---|---|
| A | The researchers estimate that the climate effects of the release of this gas could cost *$60 trillion, roughly the size of the global economy in 2012.* |
| B | Large amounts of methane are concentrated in the frozen Arctic tundra *but are also found as semi-solid gas hydrates under the sea.* |
| C | *Scientists have found plumes of gas up to a kilometer in diameter rising from these waters.* |
| D | It is thought that *up to 30% of the world's undiscovered gas and 13% of undiscovered oil lie in the waters.* |
| E | According to Lloyds of London, *investment in the Arctic could reach $100 billion within ten years.* |
| F | When you look at satellite imagery, for instance the MeToP satellite, *that's gone up significantly in the last three years and the place where the increase is happening most is over the Arctic.* |

In reading protocol, it is easy to see which stimuli pages have been seen by multiple users. In the selection menu for stimuli pages, it is annotated how many subjects have visited this page. By choosing one stimulus, the page is shown for the different users. With a tick on "show all participants on the same page" word-eye-fixations for all subjects are merged and then shown on a single stimulus page. Because reading protocol operates on the textual level, different layout versions of a stimulus can also be integrated. Therefore, within the process of merging word-eye-fixations of different users the algorithm looks, if the word cannot be found on the exact same position if it can be found in the near context of about 50 characters before or after. This way, layout variations of the same stimulus are handled more flexible and word-eye-fixations of multiple users can be integrated for one design variation. Additionally, the stimulus page can be filtered to certain AOIs and fixation data can be computed for these areas.

Showing data of multiple users on the same stimuli makes sense when they conduct the same task. We want to show examples from the search strategies- and the journalism experiment.

*Example Search strategies:* In the search strategies experiment all 32 subjects started from the same page to find related publications on the topic "education inequality". The analyst can now view these pages separated by subjects to study which page segments has been read by each user. With a click on "show all participants on the same page" the page is shown with the integrated data over all subjects. Playing around with the min fixation time slider reveals that users have read with highest fixation times the title and authors of the article (around 28sec in sum), followed by the DOI (~25sec), the tab label "References" and the category information (~23sec), then source and keywords (~15sec) and the tab label "Citations" (~12sec). This could be a starting point to compare this pattern to other pages in this experiment.

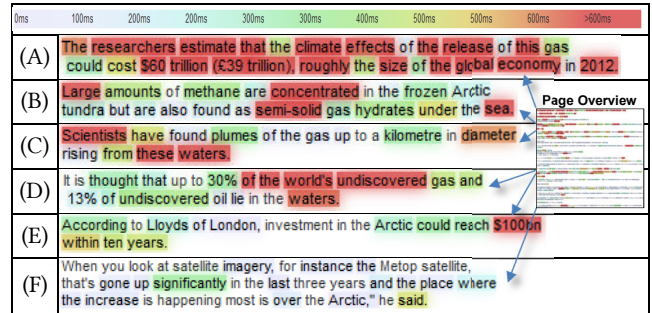

**Figure 7: In the journalism experiment participants for the Copy Editing task have found the statements A-F from Table 3 on one BBC web page. The processed reading protocol page shows that the text fragments have been read extensively with a lot of fixations over 600ms. Fixation times differ because not all participants have searched or found all statements on this page.**

*Example Journalism CPE:* In the Copy Editing task of the journalism experiment subjects were asked to verify statements from a given text by bookmarking web pages that confirm or disconfirm these statements. Table 3 shows the statements from the task description in isolation. The analyst can now check in reading protocol which web pages have been viewed by most participants in the stimuli menu. For example, the page "https://www.bbc.com/news/science-environment-23432769" has been viewed by 6 of 9 subjects. With a click, the data set is filtered to only this page showing for each subject the individual gaze behavior on it. The analyst can now check which text fragments have been read by the individual users. With a click on "show all participants on the same page" all gaze data is shown on the same page. This stimulus page is also a good example of a page that has changed its appearance in the time of the experiment because of different advertisements and news. On the textual level reading protocol can integrate the different word-eye-fixations so that we are able to see all subjects on the same stimuli. Figure 7 then shows that all verifications for this task from A) to F) can be found on the BBC web page. Showing gaze data of all subjects on the same page let the statements appear in highly red, especially the first one. This reading behavior can be found for the Copy Editing (CPE) task, but not for Story Pitch (SP) where the fixation times are higher over larger segments of the stimuli pages. Both task types have the task facets *product=find_facts* and *segment=level* in common, but differ in *goal=specific vs amorphous* and *named-items=yes vs. no* (cp. Table 1). In [14] it was found that CPE tasks have a high number of Scan-to-Read and Read-to-Scan transitions. It differs from other task types by the *level=segment* facet in opposite to *level=document*. Here we found more specifically, that for two task types with *level=segment* the one with *named-items=yes* shows a specific reading behavior with high fixation times over the single statements.

### 6.5 Further Processing

Reading Protocol provides three different mechanisms for further processing of the data. (1) A CSV table of the word-eye-fixations object can be copied for each stimulus by clicking on the button "Copy Word-Eye-Fixations as CSV to Clipboard". This table can then be used to analyze eye fixations on the word level in any other tool. (2) Processed and colored text or filtered text fragments can be easily copied from the user interface and analyzed further, e.g. by comparing read text fragments over multiple pages in isolation (3) The data table with information on task data, eye-fixations, filter for certain AOIs can be copied and pasted into any statistical software, e.g. to find correlation between certain columns.

*Example Journalism CPE:* To show an example, we will resume the example of CPE from section 6.4. We have shown that an analyst can instantly see by the hot-colored background that statements to be verified from the task have been fixated longer than the other text on the stimulus page. With the BBC web page, we already have found a good example showing this behavior. The next step would be to analyze exact fixation times for each statement and each subject.

Therefore, the analyst can copy word-eye-fixations for each subject for this stimulus from the reading protocol user interface to a statistical tool like Excel. Then word-eye-fixations for the specific statements can be separated from the rest, and by summing up a table can be generated that shows fixation times for each statement over all subjects (see Table 4). This table gives now a formal representation of the stimuli visualization that has been shown in reading protocol.

**Table 4: Fixation times in ms for each statement**

| Subject # / Statement | 13 | 32 | 61 | 66 | 7 | 83 |
|---|---|---|---|---|---|---|
| A | 2562.4 | 4219 | 5998.9 | 8312.7 | 1972.9 | 4992.8 |
| B | 2318.6 | 472.3 | 1134 | 114.7 | 2726.1 | 1050.4 |
| C | 2412.3 | 262.8 | 82.2 | 0 | 1128.2 | 988 |
| D | 5240.8 | 0 | 0 | 0 | 1120.2 | 1477.8 |
| E | 2323.3 | 0 | 0 | 0 | 743.8 | 312.6 |
| F | 723.4 | 0 | 32.7 | 0 | 49.6 | 362.6 |

The question of who has read which statement can now be answered more easily. Statement A has been read by all subjects and subject 13 has read all statements on the page. Inversely, fixation times with zero or up to 100ms indicate that the user has not read the statement.

We could go on with that analysis. However, we only want to show how word-eye-fixations can be exported and further processed in other tools to understand on a statement or text passage basis what has been read and what not, with exact fixation values.

## 7 DISCUSSION & CONCLUSION

In this section, we want to discuss the pros and cons of reading protocol and ideas for future work.

The last section showed that reading protocol provides a number of advantages for analysts of IIR experiments: (1) they can see over the whole task which pages have been scanned, which text sections have been read and for how long. (2) Viewing and reading behavior can be brought into relation with other task-data such as task types, questionnaire data and so on. (3) The whole dataset can be easily filtered to certain AOIs in order to prevent information overflow. (4) Gaze data of multiple users can be shown on the same page which helps to identify important text passages. (5) All fixation times on the word level and other task attributes can easily be exported and further analyzed in other tools. All in all, reading protocol helps to circumvent the manual inspection over dozens or hundreds of single images and videos in standard eye tracking analysis software.

However, there are also certain aspects an analyst has to be aware of when using the reading protocol software. In Web search, users are not only looking at HTML web pages but also at other media types such as images, videos, PDFs, presentations and so on. Reading protocol can only map eye tracking data to the text level when the HTML DOM is available, so other media types are ignored in the user interface. An analyst should be aware that within the search process, information read by the subjects can also come from other media types and one has to check e.g. the gaze videos or log protocols in addition.

Eye tracking data has a certain inaccuracy depending on factors such as hardware, software, calibration process and individual participant. Eye tracking hardware, for example, has a certain accuracy that is measured in degrees of visual angle, e.g. between 0.5-1 degree. One degree corresponds to a mean error of 11mm with a screen distance of 65cm. This error can even increase if the calibration process was poor; subjects are moving too much or wearing eye glasses, contact lenses or jewelry. All these factors sum up to a certain inaccuracy of the gaze data. This can be especially critical in reading situations with small-sized text [8]. This effect is the same for original eye tracking analysis software as for reading protocol; so, an analyst should take care of good gaze data quality. However, in the sections 6.4 and 6.5 we showed that eye tracking data can be good enough to recognize read statements within a web page, especially if we aggregate data over multiple users.

An aspect of future work is the aspect of learning in the search process. Vakkari [34] describes learning as changing one's knowledge structure by changes in concepts and their relations. He states: "Titles, abstracts and text passages browsed include possible ideas for restructuring her conceptual understanding of the topic". This is exactly what is processed and visualized by reading protocol. The exact fixation times are stored per single word and can be further processed per stimulus, per sentence or text passage and over the whole task. This can be a formal description of what comes to a user's mind from the system side within the search task and what can be one source of learning.

Understanding what participants have read at each point within a search task is the key to understand what might be the next action step or what they have learned. So far, with image- and video-based analysis software it is hard to track fixation times on text over the whole search process. Reading protocol automates this activity and lets the analyst understand what has been read in the search task, even over multiple participants and in specific areas of interest. In IIR experiments this will give the analyst an additional data stream that helps to understand, for example, what are the effects of the task type and task topic on reading behavior.

**Acknowledgements** This work was partly funded by the NSF grant no. IIS-1423239.